\documentclass[aps,prl,showpacs,twocolumn,amsmath,amssymb]{revtex4-1}

\usepackage{graphicx}
\usepackage{dcolumn}
\usepackage{bm}
\usepackage{units}
\usepackage{upgreek}
\usepackage{ucs}
\usepackage{layouts}
\usepackage{hyperref}
\usepackage[all]{hypcap}
\preprint{APS/123-QED}

\begin{document}

\title{Thermodynamics of Strongly Correlated One-Dimensional Bose Gases}
\author{Andreas Vogler}
\author{Ralf Labouvie}
\author{Felix Stubenrauch}
\author{Giovanni Barontini}
\author{Vera Guarrera}
\author{Herwig Ott}
\email{ott@physik.uni-kl.de}
\affiliation{Research Center OPTIMAS, Technische Universit\"at Kaiserslautern, 67663 Kaiserslautern, Germany}

\begin{abstract}
We investigate the thermodynamics of one-dimensional Bose gases in the strongly correlated regime. To this end, we prepare ensembles of independent 1D Bose gases in a two-dimensional optical lattice and perform high-resolution \textit{in situ} imaging of the column-integrated density distribution. Using an inverse Abel transformation we derive effective one-dimensional line-density profiles and compare them to exact theoretical models. The high resolution allows for a direct thermometry of the trapped ensembles. The knowledge about the temperature enables us to extract thermodynamic equations of state such as the phase-space density, the entropy per particle and the local pair correlation function.
\end{abstract}

\maketitle

One-dimensional (1D) systems exhibit a number of peculiar features which distinguish their physical behaviour from those of higher dimensional systems. The precise balance between the interaction energy, the kinetic energy and the thermal energy strongly affects their properties and give rise to a rich landscape of different regimes. Theoretically, they can be described for all temperatures and interaction strengths by exact models \cite{Lieb1963,Yang1969}, solvable with powerful numerical methods \cite{Schollwock2005} and fundamental mapping theorems \cite{Girardeau1960}. Ultracold atomic gases in one-dimensional trapping potentials provide a unique experimental platform to realize those systems, thus ideally complementing this extensive theoretical framework. 

Various physical properties of 1D Bose gases have been studied experimentally, e.g., pair correlations \cite{Kinoshita2005,  Esteve2006, Haller2009,Guarrera2012}, the momentum distribution \cite{Paredes2004, Davis2012, Jacqmin2012}, quench dynamics \cite{Cheneau2012} and full counting statistics \cite{Schumm2005}. However, an exhaustive study of the thermodynamic properties of those systems is still missing. This is of fundamental importance since the exact knowledge of thermodynamic variables and the use of precise thermometry are instrumental to study thermalization processes, which in 1D are known to be strongly reduced \cite{Kinoshita2006}. Moreover, thermodynamic quantities can be calculated for homogenous systems for all interaction strengths \cite{Yang1969} allowing for a stringent parameter-free comparison with the experimental results. Until now, these studies have been restricted to the weakly interacting regime \cite{Jacqmin2011} using magnetic microtraps \cite{Amerongen2008, Goerlitz2001, Trinker2008, Armijo2011,Jacqmin2011}. The strongly interacting regime can be reached with the help of two-dimensional optical lattices \cite{Paredes2004, Kinoshita2004, Haller2009, Guarrera2012}. However, in this case only averaged quantities have been measured such that the information on individual 1D gases is smeared out. Only recently, first temperature measurements of averaged profiles have been reported by our group \cite{Guarrera2012}. Thermodynamic studies of strongly interacting 1D gases would open new possibilities to investigate the interplay between non-equilibrium dynamics, thermalization, and strong correlations in a quantum many-body system.

In a 1D atomic gas (radial trapping frequency $\omega_r$) the effective 1D interaction strength is given by $g_{\mathrm{1D}}\approx2\hbar a \omega_r$ \cite{Olshanii1998}. Here, $a$ is the 3D scattering length. The ratio between thermal and kinetic energy \cite{Kheruntsyan2005} is the interaction parameter $\gamma$, 
\begin{equation} \label{eq:gamma}
\gamma = \frac{E_{\mathrm{int}}}{E_{\mathrm{kin}}} = \frac{m g_{\mathrm{1D}}}{\hbar^2 n_{\mathrm{1D}}}\approx \frac{2ma\omega_r}{\hbar n_{\mathrm{1D}}},
\end{equation}
where $m$ is the mass of the particle and $n_{\mathrm{1D}}$ the line-density. In the weakly correlated quasi-condensate regime the interaction parameter is $\gamma<1$ while in the strongly correlated regime it is $\gamma>1$.

Here, we study the thermodynamics of strongly correlated 1D Bose gases in a two-dimensional (2D) optical lattice. Using an inverse Abel transformation we extract effective line-density profiles from integrated \textit{in situ} density profiles. We compare the line-density profiles to exact solutions and discuss the role of temperature and interactions. Due to the inhomogeneous axial trapping potential $\omega_{\mathrm{ax}}$, the temperature is encoded in the thermal wings of each profile. This makes an \textit{in situ} thermometry possible and allows us to extract thermodynamic equations of state \cite{Ho2009,Gemelke2009,Nascimbene2010a,Nascimbene2010b,Yefsah2011,VanHoucke2012,Navon2011,Ku2012}.\\

\begin{figure}[h!]
\begin{center}
\includegraphics[width=\columnwidth]{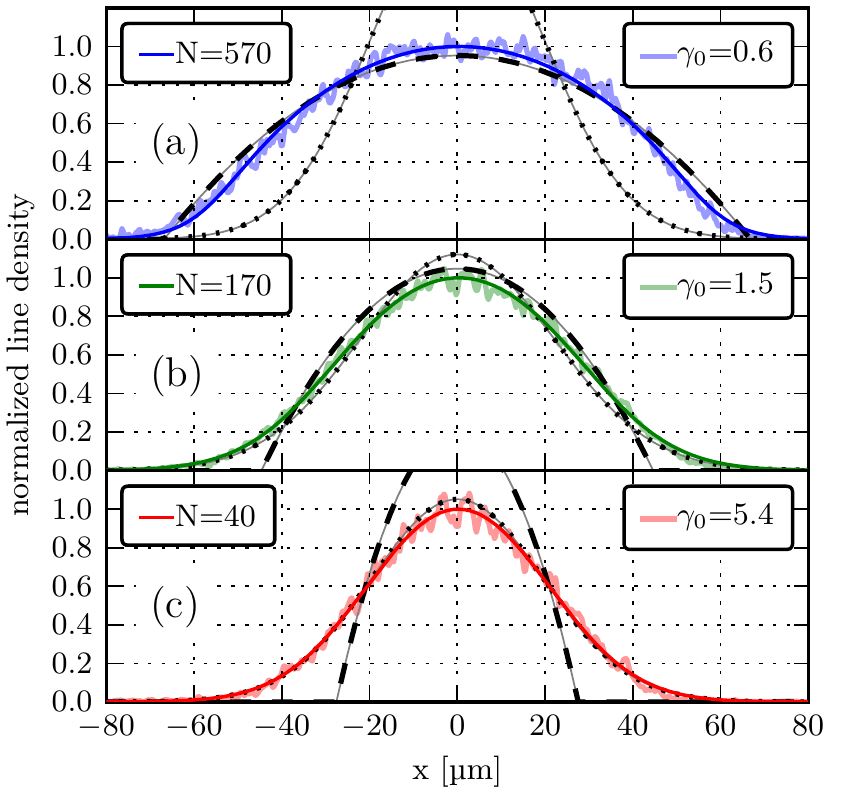}
\end{center}
\caption{(color online). (a)-(c) \textit{In situ} density profiles from data set C (light solid) for increasing distance from the trap center, corresponding to different interaction strengths and same temperature. The exact YY fits are shown as dark solid lines. $N$ denotes the number of atoms. The peak densities are $n_{\mathrm{1D}}(x=0)$: (a,b,c)=(6.2,2.7,0.7)$\times\unit[10^6]{m^{-1}}$. The asymptotic solutions with the same number of atoms as for the exact YY-fit are also shown: TF-profile $(T=\unit[0]{nK})$ (dashed) and ideal Bose gas (dotted).}
\label{fig:yycurves}
\end{figure}

We prepare a Bose-Einstein condensate (BEC) of $80\times 10^3$ ${^{87}\unit{Rb}}$ atoms, trapped in a single focused CO$_{2}$-laser beam. We use scanning electron microscopy (SEM) \cite{Gericke2008, Guarrera2011} to manipulate and probe the column-integrated density distributions. The SEM emits a focused electron beam (EB) ($I=\unit[60]{nA}$, $E=\unit[6]{keV}$, $\unit[240(10)]{nm}$ FWHM) which is moved in a rectangular pattern over the atoms. To control the number of atoms, we force an evaporation by scanning over one wing of the BEC. This additionally cools the sample to temperatures as low as $\unit[10]{nK}$. Subsequently, the BEC is loaded within $\unit[210]{ms}$ into a blue-detuned 2D optical lattice with $\lambda = \unit[774]{nm}$.

\begin{figure}[t]
\begin{center}
\includegraphics[width=\columnwidth]{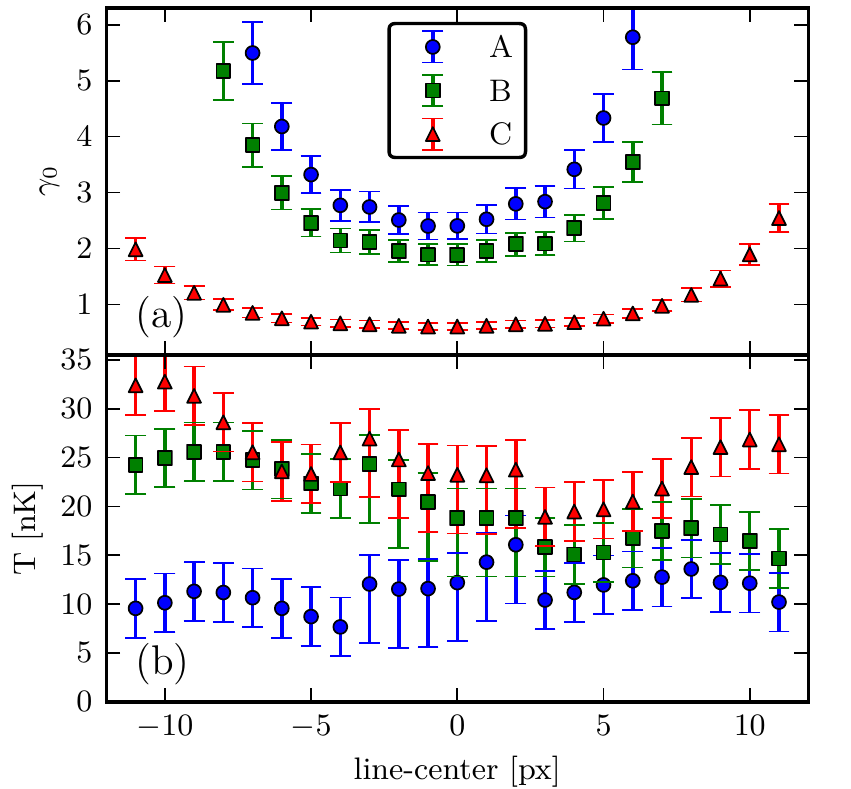}
\end{center}
\caption{(color online). (a) Central interaction strength $\gamma_0$. (b) Temperature $T$ from a fit with the exact YY theory. Every point represents one line in $R$ for the corresponding data set. Negative (positive) values of the $x$-axis represent the upper (lower) half in $R$.}
\label{fig:yyfitdata}
\end{figure}

In order to access different regimes in 1D, we prepared three samples (A,B,C) with initial atom numbers of ($10(2)$, $20(3)$ and $60(5)$) $\times 10^3$ atoms. The radial trapping frequencies in the optical lattice are $\omega_{r} = 2\pi\times \unit[56]{kHz}$ (A and B) and $\omega_{r} = 2\pi\times \unit[42]{kHz}$ (C). Simultaneously to the adiabatic ramp of the lattice, we lower the axial trapping frequency to $\omega_{\mathrm{ax}} =\unit[ 2\pi\times8.8\pm1.3]{Hz}$ (A,B) and $\omega_{\mathrm{ax}} = \unit[2\pi\times12.7\pm1.7]{Hz}$ (C). The angle between the lattice axes and the EB is $\unit[45]{\textdegree}$. 
\begin{figure}[t]
\begin{center}
\includegraphics{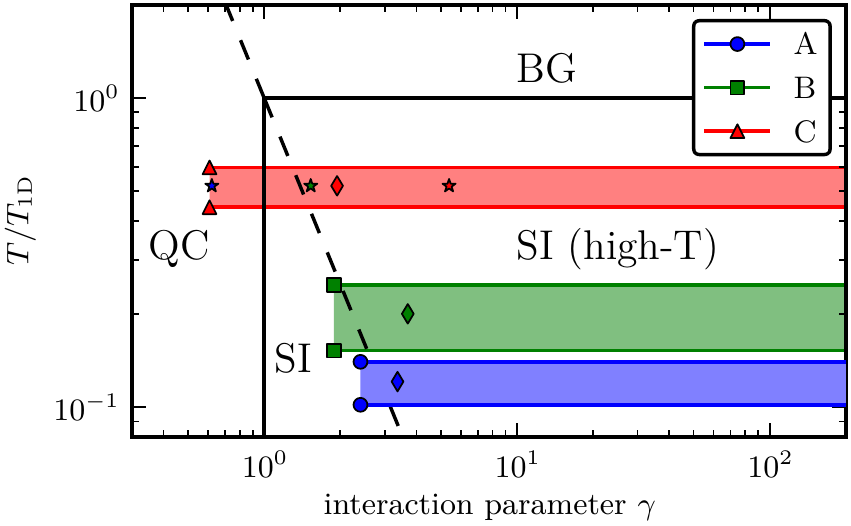}
\end{center}
\caption{(color online). Simplified phase diagram in the $\gamma-t$ plane for a uniform 1D Bose gas with repulsive contact interaction \cite{Kheruntsyan2005}. The regimes are: (SI) strongly interacting regime, (SI high-T) high-temperature regime, dominated by thermal energy, (BG) nearly ideal Bose gas and (QC) quasicondensate. The dashed line shows the degeneracy temperature $\tau=T/T_d=1$. The colored regions show the different data sets, whereas the spread in $t$ is given through the temperature error from the fit. The stars ($\star$) indicate $\gamma_0$ for the profiles shown in Fig.\ref{fig:yycurves} and the diamonds ($\Diamond$)  for the profiles in Fig.\ref{fig:tdependence}.}
\label{fig:phasediagram}
\end{figure}
Therefore, the pixel size is chosen to be $\unit[273]{nm}$ and the total imaging duration is $\unit[30]{ms}$. The fast scanning direction is oriented along the 1D gases, such that the scan speed is much faster than the speed of sound. All single-shot pictures are corrected for angle and position drifts and summed up. This yields an image sum $P$ for each of the three samples (A,B,C) containing 3200, 3900 and 1900 pictures respectively.

We decompose the integrated density profiles using an inverse Abel transformation $\mathcal{A}^{-1}$. Altough the 2D lattice in our setup has a four-fold symmetry it is smeared out due to the SEM imaging settings and the post-processing. Therefore, the prerequisite of cylindrical symmetry is approximately fulfilled. To perform the Abel inversion, we make use of the BASEX-method \cite{Dribinski2002} in a modified way \cite{basexmod}. The noisy central region \cite{abelfail} ($\pm 3$ pixel) is interpolated with a Abel inverted gaussian fit on $P$. Every horizontal line in the resulting image $R=\mathcal{A}^{-1}(P)$ corresponds to an average of all 1D gases which are at the same distance from the symmetry axis and thus have the same central interaction parameter $\gamma_0=\gamma(x\,=\,0)$. 
For every line in $R$ we perform a fit with the exact Yang-Yang theory (YY) \cite{Yang1969}, making a local density approximation \cite{Kheruntsyan2005}: $\mu(x)=\mu_0-V_{\mathrm{ax}}(x)$, where $\mu(x=0)=\mu_0$ is the central chemical potential and $V_{\mathrm{ax}}(x)=m\omega_{\mathrm{ax}}^{2}x^2/2$. The line-density is fixed by a normalization with the atom number and the pixel size, leaving the temperature as the only free parameter. As can be seen in Fig.\ref{fig:yycurves}, the fits reproduce the density profiles very well. The temperatures show only moderate variations which are compatible with the estimated error (Fig.\ref{fig:yyfitdata}b). We find (A,B,C): ($\bar{T}=\unit[11(2)]{nK}$, $\bar{T}=\unit[20(4)]{nK}$, $\bar{T}=\unit[25(4)]{nK}$). This indicates an adiabatic loading of the lattice without significant perturbations. The residual variations of $T$ originate from the inversion method as well as the interpolation in the center. The temperatures were further cross-checked via a fugacity analysis by fitting a thermal distribution to the wings of each profile. 
As shown in Fig.\ref{fig:yycurves}, the density profiles change drastically with the interaction parameter $\gamma_0$. This is due to the reduction of interaction energy $E_{int}\simeq n_{\mathrm{1D}} g_{\mathrm{1D}}$. Note, that even though $\gamma_0$ is increasing towards the outer tubes, the absolute value of the interaction energy drops as $n_{\mathrm{1D}}$. The critical density at which the thermal energy dominates is defined via the dimensionless degeneracy temperature $\tau(x)=T/T_d(x)$ with $T_d(x)=\hbar^2 n(x)^2 / 2m$ \cite{Kheruntsyan2005}. For Fig.\ref{fig:yycurves}a the value in the center is $\tau(0)=0.2$ and the density profile is close to a Thomas-Fermi distribution. For the high-temperature region ($\tau(0)=15$, Fig.\ref{fig:yycurves}c) the effect of interaction is masked, because the mean inter-particle distance is larger than the thermal de-Broglie wavelength and the system is dominated by the thermal energy, resulting in a thermal distribution. The border ($\tau=1.2$, Fig.\ref{fig:yycurves}b) between the two regimes is not sharp and as a consequence, the increasing effect of thermal energy towards the wings is smooth.
\begin{figure}[t!]
\begin{center}
\includegraphics[width=\columnwidth]{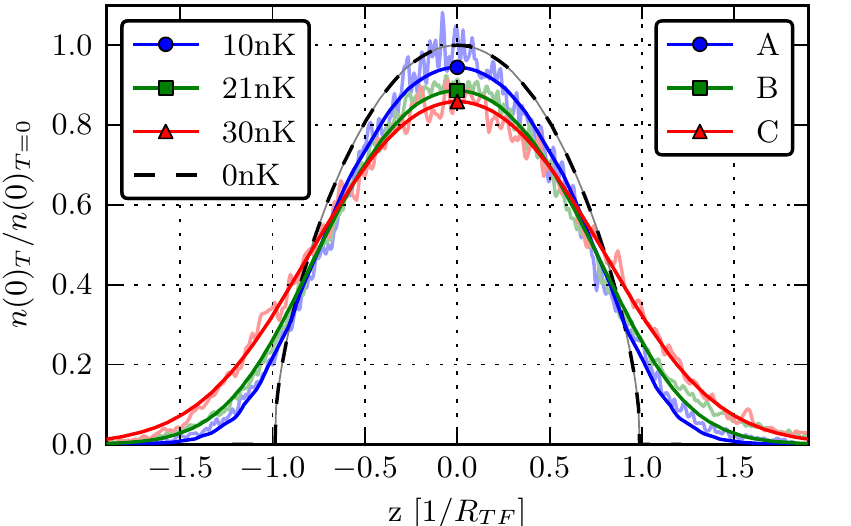}
\end{center}
\caption{(color online). Comparison of selected density profiles from all three data sets with same atom number $N=120(10)$ and different temperatures with the theoretical prediction of the YY theory at $T=\unit[0]{nK}$. The atoms agglomerate in the center for lower temperatures.}
\label{fig:tdependence}
\end{figure}
To highlight the role of the temperature in the strongly interacting regime it is convenient to normalize T to a density-independent energy scale $T_{\mathrm{1D}} = mg_{\mathrm{1D}}^2 / 2\hbar^2 k_{\mathrm{B}}$, where $k_{\mathrm{B}}$ is Boltzmann's constant \cite{Kheruntsyan2005}. To enter the strongly interacting regime (SI), the dimensionless temperature $t$ has to fulfill $t=T/T_{\mathrm{1D}}<1$, which is the case for all data sets. Using the dimensionless parameters $\gamma$ and $t$, we can depict a phase diagram for a uniform 1D Bose gas \cite{Kheruntsyan2005}, as shown in Fig.\ref{fig:phasediagram} together with the experimentally accessed regimes. Data set C covers all regimes that are below $t=1$, from the weakly correlated quasi-condensate regime (QC), crossing the SI regime to the high-temperature SI regime. To  visualize the effect of the temperature, we now compare density profiles with different temperatures but with the same number of atoms $N=120(10)$ (Fig.\ref{fig:tdependence}). In order to account for different axial trapping frequencies, the axial coordinate $x$ is normalized by $R_{TF}=\left(\frac{3Ng}{m\omega_{\mathrm{ax}}}\right)^{1/3}$\cite{Kheruntsyan2005}. For comparison, a YY profile at $T=0$ is shown.
\begin{figure*}[ht!]
\begin{center}
\includegraphics{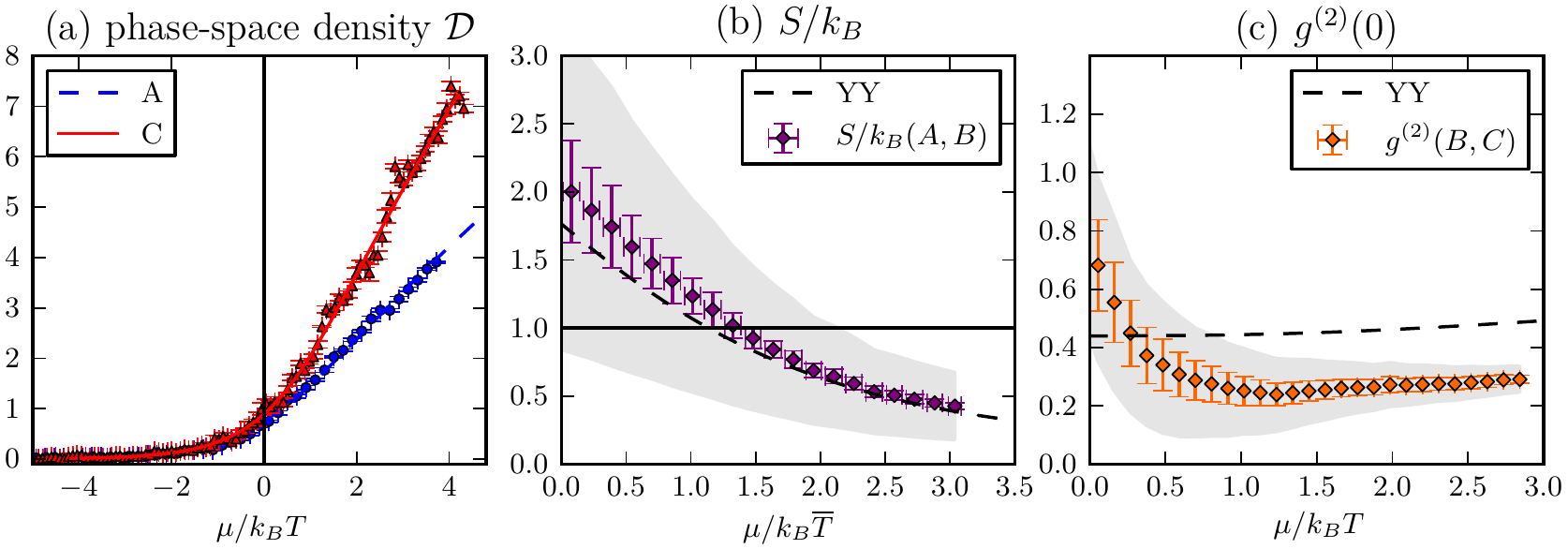}
\caption{(color online). (a) Phase space density $\mathcal{D}$ for data set A and C together with the exact YY prediction (dashed/solid). (b) Entropy per particle $S/k_B$ with YY prediction (dashed) and calculated with set A and B (equal $g_{\mathrm{1D}}$). The shaded area shows the error including the temperature uncertainty. (c) Local pair correlation function $g^{(2)}(0)$ calculated from sets B and C (similar $T$) with YY prediction (dashed). The shaded area shows the result for different trapping frequencies in the range of $\omega_{\mathrm{ax}}=2\pi\times(12.7\pm0.05)$Hz.} 
\end{center}
\label{fig:EOS}
\end{figure*}
With decreasing temperature, the density enhances in the center and converges to the $T=0$ prediction. 

Finally, our method allows for the extraction of thermodynamical quantities. Every pixel in $R$ is an independent measurement of $n_{\mathrm{1D}}(g, \mu(x), T)$, yielding access to thermodynamic equations of state \cite{Ho2009}. The phase space density $\mathcal{D}=\lambda_{\mathrm{dB}}n_{\mathrm{1D}}$ follows as a direct result from $R$ (Fig.\ref{fig:EOS}a). Additionally, integrating $n_{\mathrm{1D}}(g, \mu(x), T)$ yields the local pressure of the system
\begin{equation}
P(g_{\mathrm{1D}},\mu,T)=\int_{- \infty}^{\mu}n_{\mathrm{1D}}(g_{\mathrm{1D}},\mu',T)d\mu',
\label{eq:P}
\end{equation}
where the integral can be approximated by a discrete sum. For a system in local thermal equilibrium, the Gibbs-Duhem relation $dP=nd\mu+sdT$ is applicable, where $s$ is the entropy density. Using eq.(\ref{eq:P}), the entropy per particle $S$ reads:
\begin{equation}
S=\frac{1}{n_{\mathrm{1D}}(g_{\mathrm{1D}},\mu,T)}\left( \frac{\partial P(g_{\mathrm{1D}},\mu,T)}{\partial T}\right)_{g_{\mathrm{1D}},\mu},
\label{eq:S}
\end{equation}
where the derivative is taken for fixed values of $g_{\mathrm{1D}}$ and $\mu$. To compute the derivative in eq.(\ref{eq:S}), we use the local pressures obtained from sets A and B. The derivative can be approximated with the finite difference quotient $(P_A-P_B) / (T_A-T_B)$ and $n_{\mathrm{1D}}(\mu)$ as the average density. The result is shown in Fig.\ref{fig:EOS}b together with the prediction of the YY theory for an intermediate temperature of $\unit[18]{nK}$. The minimum value of $0.4\times k_B$ is comparable to the entropy per particle for a 2D Mott insulator of $~0.3\times k_B$, reported in \cite{Sherson2010}. A lower value of $~0.06\times k_B$, but for a weakly interacting 2D Bose gas was reported in \cite{Yefsah2011}. The agreement with the Yang-Yang theory is remarkable for large values of $\mu$, while for smaller values, the spread in temperature does not allow for a precise determination. Another quantity of interest in a strongly interacting system is the local pair-correlation function $g^{(2)}(0)$. For many-body systems with repulsive contact interaction, the interaction Hamiltonian reads $\hat{H}_\mathrm{int}=\frac{g_{\mathrm{1D}}}{2}\int \hat{\Psi}^{\dagger}({\bf x})\hat{\Psi}^{\dagger}({\bf x}) \hat{\Psi}({\bf x}) \hat{\Psi}({\bf x}) d{\bf x}$, where $\hat{\Psi}({\bf x})$ and $\hat{\Psi}{\dagger}({\bf x})$ are the field operators. Using the Hellmann-Feynman theorem \cite{Hellmann1933, Feynman1939}, one can show \cite{Kheruntsyan2005} that $g^{(2)}(0)$ in a homogeneous system is then given by
\begin{equation}
g^{(2)}(0)=\frac{\left\langle \hat{\Psi}^{\dagger}({\bf x})^2 \hat{\Psi}({\bf x})^2 \right\rangle} {\left\langle \hat{\Psi}^{\dagger}({\bf x}) \hat{\Psi}({\bf x}) \right\rangle ^2}=-\frac{2}{n_{\mathrm{1D}}^2}\left( \frac{\partial P(g_{\mathrm{1D}},\mu,T)}{\partial g_{\mathrm{1D}}} \right)_{T,\mu}.
\label{eq:g2}
\end{equation}
We approximate eq.(\ref{eq:g2}) by using $P(g_{\mathrm{1D}},\mu,T)$ of set B and C. The results are shown in Fig.\ref{fig:EOS}c. The measured $g^{(2)}(0)$ underestimates the predicted result of the YY model by 50\%. This is due to the small relative difference between the two pressures of less than 10\% which makes the procedure very sensitive to little deviations from the ideal density profile and amplifies the influence of the axial trap frequency and the finite temperature difference. Nevertheless, the minimal value of $g^{(2)}(0)\approx0.3$ clearly indicates strong anti-bunching for a large range of values of the chemical potential -- as expected for a partially fermionized system.

In conclusion, we have measured effective \textit{in situ} density profiles of strongly correlated 1D Bose gases. The density profiles show excellent agreement with the Yang-Yang thermodynamic theory. The high precision allows for a direct determination of the temperature of the atoms in the two-dimensional optical lattice. We discuss the role of temperature and its interplay with the atomic interactions and derive several thermodynamic equations of state: the phase space density, the entropy per particle and the local pair-correlation function. Our approach paves the way to study \textit{in situ} thermalization processes and dynamical properties of ultracold atomic gases in optical lattices as well as in bulk systems.

\begin{acknowledgments}
We thank P. W\"urtz and D. Muth for technical support and helpful discussions. We acknowledge financial support by the DFG within the SFB/TRR 49, the GRK 792, and the MAINZ graduate school. V.G. and G.B. are supported by Marie Curie Intra-European Fellowships. 
\end{acknowledgments}

\bibliographystyle{apsrev4-1}

\begin{references}
\bibitem{basexmod}
We use the BASEX method in combination with a basis-set of radially centered gaussians of the form $f_k(r) = exp(-\frac{-r^2}{2 k^2})$. The basis-functions are invariant under Abel inversions: $A^{-1}(f_k(r)) = (k\sqrt{2\pi})^{-1}f_k(r)$. We evaluate the upper and lower half of P individually in order to account for possible heating mechanisms during the finite imaging time.
\bibitem{abelfail}
The information that is contained in the column-integrated picture $P$ about a region within a certain radius $r$ scales with $r$. The closer to the symmetry axis, the less information about this region can be obtained. In particular, the pixels on the symmetry axis contain the only information about the center-line. In combination with noise this leads to a failure of Abel inversion methods close to the symmetry axis.
\bibitem{oscfreq}
The change in the transverse oscillation frequency (which is needed to change $g$) affects the axial confinement. This makes a separate determination of the axial oscillation frequency for different interaction strengths in our system inavoidable.
\end{references}

\end{document}